\documentclass[a4paper]{article}

\usepackage{INTERSPEECH2020}
\usepackage{amsmath}
\usepackage{xcolor}
\usepackage{subcaption}

\title{A learned conditional prior for the VAE acoustic space of a TTS system}
\name{authors}
\name{Penny Karanasou, Sri Karlapati, Alexis Moinet, Arnaud Joly, Ammar Abbas, Simon Slangen, Jaime Lorenzo Trueba, Thomas Drugman}
\address{
Amazon Research, Cambridge, United Kingdom}
\email{\{pkarana, srikarla, amoinet, jarnaud, syeabbs, slangens, truebaj, drugman\}@amazon.com}

\begin{document}

\maketitle
\begin{abstract}
Many factors influence speech yielding different renditions of a given sentence. Generative models, such as variational autoencoders (VAEs), capture this variability and allow multiple renditions of the same sentence via sampling. The degree of prosodic variability depends heavily on the prior that is used when sampling. In this paper, we propose a novel method to compute an informative prior for the VAE latent space of a neural text-to-speech (TTS) system. By doing so, we aim to sample with more prosodic variability, while gaining controllability over the latent space's structure.

By using as prior the posterior distribution of a secondary VAE, which we condition on a speaker vector, we can sample from the primary VAE taking explicitly the conditioning into account and resulting in samples from a specific region of the latent space for each condition (i.e. speaker). 
A formal preference test demonstrates significant preference of the proposed approach over standard Conditional VAE. We also provide visualisations of the latent space where well-separated condition-specific clusters appear, as well as ablation studies to better understand the behaviour of the system.
\end{abstract}
\noindent\textbf{Index Terms}: Learned Conditional Prior for VAE, disentanglement, controllable TTS,  multi-speaker model, acoustic latent space, Conditional VAE, neural TTS

\section{Introduction}

Numerous factors, such as linguistic content, speaking style, dialect, speaker identity, emotional state and environment, influence speech and make variability one of its inherent characteristics. The presence of one or many of these factors results in different renditions of a given sentence. However, typical TTS systems are unable to produce multiple renditions of the same sentence. That is why recently probabilistic generative models, such as VAEs, have been used to learn a latent representation that captures the variability of speech \cite{ChengHsu2016, Hsu2017, Akuzawa2018, Zhang2019}. This latent space captures a distribution over multiple renditions and, thus, sampling from it can produce varied speech.

In the standard approach, sampling is still done from the VAE latent space by using a standard Gaussian prior. Such a prior, even if it still represents natural prosody, loses some of the variability of real speech, and has been shown to lead to over-regularization and poor latent representations \cite{Dilokthanakul2016, Bowman2016}. In addition, one has no controllability on the part of the latent space from which sampling is achieved. 

In this paper, we introduce a more informative prior that handles the aforementioned caveats. This prior is conditioned on a specific factor that influences speech and introduces condition-specific clusters to the VAE space. In that way, the variability induced by a given condition is kept in the latent space. Then, by providing the desired conditioning during inference, we guarantee sampling from a specific part of the latent space which captures the condition identity and prosodic variability. In other words, we achieve controllability of the generated speech thanks to the disentanglement that results from using a conditional prior.

More precisely, we suggest to adopt a learned conditional posterior as prior, following the approach of \cite{Aliakbarian2019}. We introduce a hierarchical VAE structure with a secondary VAE which takes as input a condition influencing speech, in our case a speaker vector, and learns the distribution of this conditioning signal. This can also be seen as generating TTS-specific speaker embeddings that are trained jointly with the rest of the system. The learned posterior distribution then acts as the prior of the primary encoder. This is a novel method to use a learned conditional prior in TTS. 

The typical approach to add a condition to a VAE is the Conditional VAE (CVAE) \cite{Akuzawa2018, Skerry-Ryan2018}. In CVAE, the condition is concatenated to the VAE encoder and decoder inputs. This maintains the identity of the condition, while sampling is done in practice from a standard Gaussian prior. In other words, although the encoder and decoder are conditional, the latent space itself, as expressed by the prior, isn't. In our approach, in contrast, sampling is done from a learned conditional prior. This gives us controllability over the sampling from a structured latent space. In addition, using a more informative prior generates samples with more prosodic variability. 

There has also been work recently on learning more potent latent representations by specifying more complex priors over the VAE latent space. VampPrior \cite{Tomczak2018, Hodari2020} is an example of such approaches where the proposed prior is a mixture of variational posteriors conditioned on learnable pseudo-data. This is close to our approach from the perspective that the parameters of the prior are learned jointly with the rest of the model. In our case, however, the input is the actual conditioning, so there is no need to define pseudo-units which is not always an obvious task. Another related work is the use of Gaussian mixture as a latent prior \cite{Hsu2019}  where there is a different speech factor represented in each mixture component. This work focuses on the disentanglement of different factors that influence the latent space.

In controllable TTS literature, in \cite{Wang2018} the authors introduce the ``global style tokens'' (GST), style embeddings that condition the text encoder and are learned jointly with Tacotron \cite{Wang2017}. This architecture is not variational though, thus with no direct way to sample utterances of varying prosody. Other works adopt a variational method and condition the VAE on contextual linguistic information, either directly \cite{Tyagi2020} or via training a prediction model that is then applied to sampling \cite{Karlapati2020, Hodari2021}.  These works are related to ours from the perspective of learning TTS-specific embeddings jointly with the rest of the system. 

Another direction of gaining controllability and interpretability on the latent space is by introducing a hierarchical linguistic structure \cite{Sun2020, Hono2020}, where these fine-grained latent variables can control different speech factors. Our method can be seen as complementary to these approaches, with the possibility to build a hierarchical structure on top of it. 

The main contribution of this paper is that it introduces a more informative prior to the acoustic latent space, jointly trained with the rest of the system in a hierarchical VAE architecture. Using this prior adds to the expressivity of the generated samples compared to a CVAE baseline system, while maintaining the input condition. It also provides controllability on the sampling from a structured VAE latent space. While this method is successfully tested on a multi-speaker model, we hypothesize that such a prior could be learned on any condition that influences speech, e.g. language, style, emotion, etc.

\section{Models}

\subsection{Variational Autoencoder (VAE) and Conditional VAE (CVAE)}
\label{sec:vae}

A VAE \cite{Kingma2014} is a generative model where a set of latent variables $z$ is sampled from the prior distribution $p_\theta(z)$ and the data $x$ is generated by the generative distribution $p_\theta(x|z)$ conditioned on $z$, where $\theta$ is the set of parameters of the model. It is trained to maximize the log-likelihood $p_\theta (x)$ by using the evidence lower bound (ELBO) of the log-likelihood as the objective function:
\begin{equation}
ELBO(\theta,\phi) = -KL (q_\phi(z|x) \| p_\theta(z)) +  \mathbb{E}_{q_\phi(z|x)}[\log p_\theta(x|z)],
\end{equation}  
where $q_\phi(z|x)$ is an introduced approximate posterior to address the intractability of the true posterior $p_\theta(z|x)$ in maximum likelihood inference. It can be seen as a regularized version of an autoencoder, where $q_\phi(z|x)$ can be considered as the encoder and $p_\theta(x|z)$ as the decoder. This objective consists of two terms. The first term is the Kullback-Leibler (KL) divergence that tries to make the approximate posterior $q_\phi(z|x)$ close to the prior $p_\theta(z)$. The second term is the reconstruction loss that tries to make the reconstructed data close to the original ones. 

The prior over latent variables $p_\theta(z)$ is assumed to follow a standard Gaussian distribution $\mathcal{N}(z;0,\mathbb{I})$, where $\mathbb{I}$ is the identity matrix. The approximate posterior $q_\phi(z|x)$ is also modelled by a Gaussian distribution $\mathcal{N}(z;\mu,\sigma^2)$. Following the ``reparametrization trick'' \cite{Kingma2014}, sampling $z$ from distribution  $\mathcal{N}(z;\mu,\sigma^2)$ is decomposed to first sampling from $\epsilon \sim \mathcal{N}(z;0,\mathbb{I})$ and then computing 
\begin{equation}
z=\mu+\sigma \odot \epsilon .
\label{eq:reparam_trick}
\end{equation}

CVAE \cite{Sohn2015} is an extension of VAE, where the distribution of the output space $x$ is modelled by a generative model conditioned on the conditioning variable or observation $c$. The model is now trained to maximize the conditional log-likelihood, $\log p_\theta(x|c)$, thus the variational training objective becomes
\begin{multline}
ELBO(\theta,\phi) =-KL (q_\phi(z|x,c) \| p_\theta(z|c)) +\\
 \mathbb{E}_{q_\phi(z|x,c)}[\log p_\theta(x|z,c)],
\end{multline}
where both the encoder $q_\phi(z|x,c)$ and the decoder $p_\theta(x|z,c)$ are conditioned on the conditioning variable $c$. Also, the prior distribution $p_\theta(z|c)$ is conditioned on $c$. In practice, however, the prior distribution of the latent variable is assumed to be independent of $c$, i.e. $p_\theta(z | c) = p_\theta(z)$. This ignores the conditioning in the sampling process.

\subsection{Seq2seq TTS model with CVAE}
\label{sec:seq2seq}

\begin{figure}[ht]
\centering
\includegraphics[width=\linewidth]{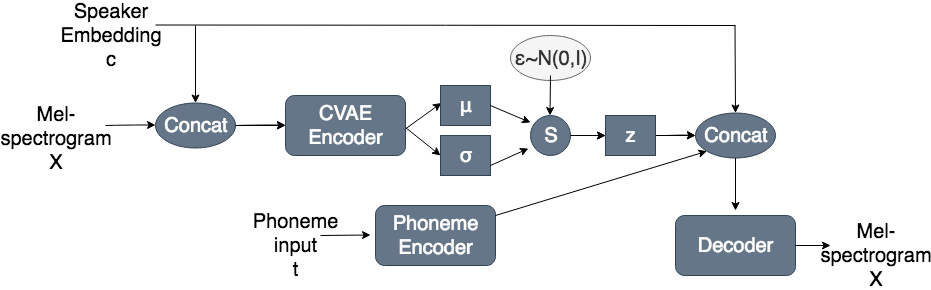}
\caption{CVAE architecture for TTS. The given condition, speaker embedding, is concatenated to the inputs of the VAE encoder and of the decoder.}
\label{fig:cvae}
\end{figure}

As can be seen in Figure \ref{fig:cvae}, our baseline network is a sequence-to-sequence (seq2seq) TTS model that converts an input phoneme sequence to a generated mel-spectrogram. The input to the decoder is the concatenation of the states of two encoders: a phoneme encoder that encodes an input phoneme sequence, and a reference encoder that encodes a reference audio sequence (mel-spectrogram) into a fixed-length low-dimensional latent representation. The architecture of the two encoders is the same as in Tacotron2 with a VAE \cite{Zhang2019}. The prior and approximative posterior of the VAE are the Gaussian distributions mentioned in section \ref{sec:vae}. The location-sensitive attention module and autoregressive decoder have the same architecture as Tacotron2 \cite{Shen2018}. Lastly, a parallel WaveNet vocoder \cite{Oord2018} is utilized to reconstruct waveforms. 

In Figure \ref{fig:cvae} an extra input source, ``Speaker embeddings'', is concatenated to the input of the VAE encoder and to the input of the decoder. This transforms our reference encoder to a CVAE where the condition $\vec{c}$ is an embedding that represents the speaker identity. The speaker identity is maintained during inference by conditioning the decoder on this signal and concatenating with the samples from the acoustic latent space. This is similar to \cite{Akuzawa2018} where they combine a CVAE with an autoregressive model. In the latter, though, the VAE is conditioned on text, while in our case on speaker embeddings.

In our system the ELBO is:
\begin{multline}
ELBO(\theta,\phi)  = -\lambda KL (q_\phi(\vec{z}|\vec{X},\vec{c}) \| p_\theta(\vec{z})) +\\
  \mathbb{E}_{q_\phi(\vec{z}|\vec{X},\vec{c})}[\log p_\theta(\vec{X}|\vec{z},\vec{t},\vec{c})],
\label{eq:loss}
\end{multline}
where the reconstruction loss term is now conditioned on the latent variable $\vec{z}$, the input text $\vec{t}$ and the condition $\vec{c}$. To avoid KL collapse, the KL annealing scheme of \cite{Bowman2016} is used and the KL term is weighted by the annealing weight $\lambda$.

\subsection{Learned Conditional Prior VAE (LCPVAE)}
\label{sec:cpvae}

We present LCPVAE, our proposed architecture, in Figure 2. It is a seq2seq model that consists of two VAE encoders, the Conditional Primary and Secondary VAE encoders (CPVAE and CSVAE respectively). For the dark blue-coloured part of the network, the architecture described in Section \ref{sec:seq2seq} is maintained. The light grey-coloured part of the network, the CSVAE encoder and decoder, consist of simple feedforward layers. The Primary VAE encoder is the ``Conditional Primary VAE'' (CPVAE) encoder with mel-spectrogram as input. The Secondary VAE encoder is the ``Conditioning Signal VAE'' (CSVAE) that takes as input the condition, in our case speaker vector, and learns the distribution of this condition. The posterior distribution from CSVAE then acts as the prior to the sampler of the CPVAE encoder. Thus, we explicitly make the sampling of the latent variable depend on the condition. In other words, instead of using $p_\theta(\vec{z})$ as prior distribution, we truly use a learned posterior $p_\theta(\vec{z}|\vec{c})$. 

\begin{figure}[ht]
\centering
\includegraphics[width=\linewidth]{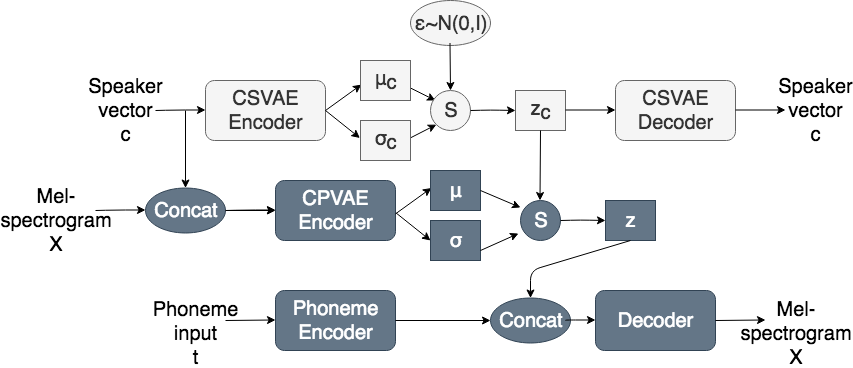}
\caption{LCPVAE architecture for TTS. A secondary VAE, CSVAE (light grey-coloured), learns a distribution over a given input condition, speaker vector. Then we use this learned prior to sample from the primary VAE, CPVAE.}
\label{fig:cpvae}
\end{figure}

As can be seen in Figure \ref{fig:cpvae} we still condition the CPVAE via concatenation, since the two inputs to the encoder, i.e., the data (mel-spectrogram) and the condition (speaker vector), are deterministic and useful to compress the sample into the latent space. The decoder is conditioned via the hierarchical VAE described above.

We define the prior of CSVAE to be a standard Gaussian distribution $\mathcal{N}(\vec{z}_c;\vec{0},\mathbb{I})$, same as in standard VAE described in Section \ref{sec:vae}, and we sample following the reparametrization trick of Eq. \ref{eq:reparam_trick}. Concerning CPVAE, we take its posterior not directly normally distributed but conditioned on the posterior of CSVAE. To this end, we adopt the extended reparametrisation trick from \cite{Aliakbarian2019} to sample from the CPVAE space:
\begin{eqnarray}
\vec{z} & = &\vec{\mu} + \vec{\sigma} \odot \vec{z}_c \nonumber \\
  & = &\vec{\mu} + \vec{\sigma} \odot (\vec{\mu}_c + \vec{\sigma}_c \odot \epsilon) \nonumber \\
  & = & (\vec{\mu} + \vec{\sigma} \odot \vec{\mu}_c) + (\vec{\sigma} \odot \vec{\sigma}_c)\odot \epsilon,
 \label{extended_param_trick}
\end{eqnarray}
where $\vec{\mu}_c$ and $\vec{\sigma}_c$  are generated from the CSVAE encoder and $\vec{z}_c$ is a sample from the CSVAE latent space. $\vec{\mu}$ and $\vec{\sigma}$ are generated from the CPVAE encoder  and $\vec{z}$ is a sample from the CPVAE latent space. While $\vec{z}_c$ follows the reparametrization trick of Eq. \ref{eq:reparam_trick} where $\mathcal{N}(\vec{z}_c; \vec{0},\mathbb{I})$ acts as the prior, the extended reparametrization trick samples $\vec{z}$ using the posterior of CSVAE $\mathcal{N} (\vec{z}; \vec{\mu}_c, \vec{\sigma}_c^2)$ as prior. 

During inference, in CSVAE the samples are drawn from the posterior since the conditioning is still provided as one-hot speaker vectors. Then, the LCPVAE decoder samples a latent variable from the posterior of CSVAE given the condition (rather than a general prior distribution) and generates a sample. This results in samples from a very similar region if conditioned on the same signal. 

The complete loss of the system is:
\begin{equation}
\mathcal{L} = \lambda(\mathcal{L}^{\it{CSVAE}}_{KL} + \mathcal{L}^{\it{CPVAE}}_{KL}) + \mathcal{L}^{\it{CSVAE}}_{rec} + \mathcal{L}^{\it{CPVAE}}_{rec},
\label{eq:loss_lcpvae}
\end{equation}
where 
\begin{equation}
\mathcal{L}^{\it{CSVAE}}_{KL} = KL (\mathcal{N} (\vec{\mu}_c, \vec{\sigma}_c^2) \| \mathcal{N}(\vec{0},\mathbb{I}))
\label{eq:kl_csvae}
\end{equation}
and
\begin{equation}
\mathcal{L}^{\it{CPVAE}}_{KL} = KL( \mathcal{N}(\vec{\mu} + \vec{\sigma} \odot \vec{\mu}_c, (\vec{\sigma} \odot \vec{\sigma}_c)^2 ) \| \mathcal{N} (\vec{\mu_c}, \vec{\sigma}_c^2))
\label{eq:kl_cpvae}
\end{equation}

We have two KL terms, one for CSVAE and one for CPVAE. The CSVAE one is the standard one for a VAE encoder defined as the KL divergence between its posterior and the standard Gaussian prior (Eq. \ref{eq:kl_csvae}). In the case of CPVAE, its posterior is now a Gaussian distribution $\mathcal{N}(\vec{\mu} + \vec{\sigma} \odot \vec{\mu}_c, (\vec{\sigma} \odot \vec{\sigma}_c)^2 )$. This modifies the CPVAE KL loss to the KL divergence between the posterior of CPVAE and the posterior of CSVAE (Eq. \ref{eq:kl_cpvae}). We freeze the weights of CSVAE before computing the CPVAE KL divergence to keep the posterior of the CSVAE stable and move the posterior of the CPVAE. The same KL annealing scheme as in baseline model is used both for CSVAE and CPVAE KL losses. On top of the reconstruction loss of the Primary VAE ($\mathcal{L}^{\it{CPVAE}}_{rec}$ in Eq. \ref{eq:loss_lcpvae}), we add an L1 loss as the reconstruction loss of the Secondary VAE ($\mathcal{L}^{\it{CSVAE}}_{rec}$ in Eq. \ref{eq:loss_lcpvae}).

\section{Experiments}

\subsection{Data}

Experiments were conducted on an internal  American English conversational data corpus of 12 speakers, 6 male and 6 female, with approximately 5 hours of speech per speaker. All speakers are normalised to 1 channel, 24 kHz sampling rate and 16 bit signed-integer PCM audio. All normalised audio is processed to trim silences longer than 50 ms. The data are split into training, validation and test sets of approximately 57000, 600 and 600 utterances respectively.  

The phoneme encoder takes as input phoneme features consisting of one-hot phone identity, including service tokens such as word-boundaries. 
The input to CVAE and CPVAE encoders are 80 mel-band, 50Hz-12kHz mel-spectrograms with 12.5ms frame-shift.

CVAE is conditioned on 192-dimensional speaker embeddings, inferred on 24 kHz mel spectrograms. They represent the speaker embedding associated with the input utterance and are averaged per speaker. These embeddings were trained with a Generalised end-to-end (GE2E) network for speaker verification \cite{Wan2018}. We also experimented with using one-hot speaker vectors instead of pretrained embeddings as input to CVAE. 
The CVAE with pretrained speaker embeddings as input was empirically found to outperform the CVAE with one-hot speaker vectors as input. Thus, it was decided to keep it as our baseline.

\subsection{Evaluation}
\label{sec:eval}

\begin{figure}[ht]
\centering
\includegraphics[width=\linewidth]{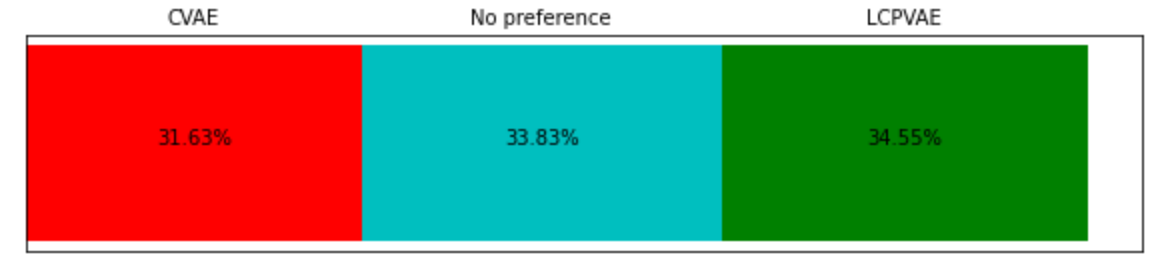}
\caption{Preference test: LCPVAE is preferred over baseline CVAE in terms of expressivity with a p-value of 0.006}
\label{fig/preference_test_cpvae}
\end{figure}

A preference test was conducted in Amazon Mechanical Turk comparing LCPVAE against CVAE. The evaluation set was made up of 25 utterances per speaker, including all male and female speakers of the test set. It was required to have 25 listeners per utterance. The question asked was which system had more expressive speech, and LCPVAE was preferred over CVAE with a p-value of 0.006. This is a significant preference for the herein proposed approach. Figure \ref{fig/preference_test_cpvae} is a graphical representation of the preference test results.

\begin{figure*}[ht]
\centering
\begin{minipage}[b]{.3\textwidth}
\includegraphics[width=\linewidth]{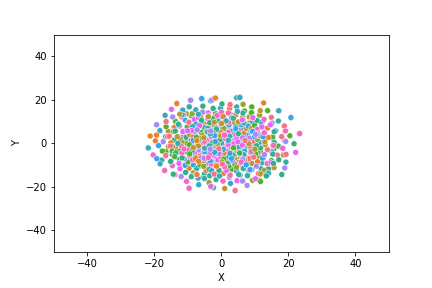}
\subcaption{CVAE latent space}
\label{fig:visualisations-cvae}
\end{minipage}\qquad
\begin{minipage}[b]{.3\textwidth}
\includegraphics[width=\linewidth]{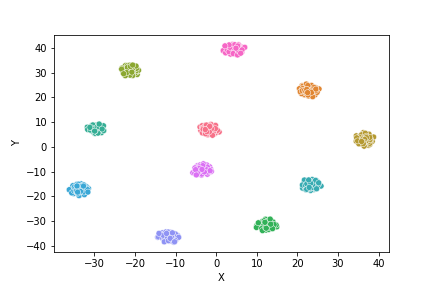}
\subcaption{CSVAE latent space}
\label{fig:visualisations-csvae}
\end{minipage}\qquad
\begin{minipage}[b]{.3\textwidth}
\includegraphics[width=\linewidth]{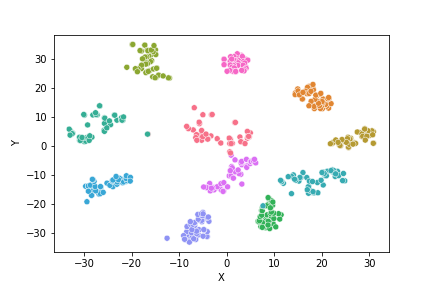}
\subcaption{CPVAE latent space}
\label{fig:visualisations-lcpvae}
\end{minipage}
\caption{t-SNE visualisations of latent spaces, where each speaker is represented with a different colour. The CVAE space is seen to the left (plot a), where no structure is observed. In the middle (plot b) the CSVAE space is shown. Training the CSVAE encoder to reconstruct the speaker vector results in a low dimensional speaker space with well separated clusters. The CPVAE latent variables (plot c) are then shifted based on the learned posterior of the CSVAE space and clusters appear in this space too.}
\vspace{-0.1cm}
\label{fig:visualisations}
\end{figure*}

To further interpret the results, we conducted visualisations of the latent spaces of CVAE, CSVAE and CPVAE, presented in Figure \ref{fig:visualisations}. We used the t-distributed stochastic neighbour embedding (t-SNE) \cite{vanDerMaaten2008} visualisation algorithm. Each colour corresponds to a different speaker and each dot to a sample. 

First, in Figure \ref{fig:visualisations-cvae} the t-SNE visualisation of the CVAE space is presented. No clusters are present and the samples are aggregated around zero. This is expected as the samples are drawn from a standard Gaussian prior. By providing the speaker vector as input to both the encoder and the decoder, this condition is factored out from the latent space leaving an unstructured space centered around the zero mean of the prior. During inference, the decoder samples latent variables from this space and concatenates them with the conditioning signal. This results in samples that all respect the provided condition. However, sampling from a standard Gaussian prior might limit the captured prosodic variability.

The behaviour of the CPVAE latent space is very different. In Figure \ref{fig:visualisations-lcpvae}, clearly separated speaker clusters appear. At inference time, contrary to CVAE, the LCPVAE decoder does not sample from a general prior distribution, but from a posterior given a condition. This is made possible because we can use the CSVAE encoder to approximate the posterior of each condition. In Figure \ref{fig:visualisations-csvae}, the CSVAE space manages to produce clean and dense speaker clusters, which results in a good approximation of the posterior given a certain speaker. The CSVAE latent space can also be interpreted as learning task-specific speaker embeddings, since low-dimensional clustered speaker representations appear there. Then, the CPVAE latent variables are shifted following the extended reparametrization trick of Eq. \ref{extended_param_trick} and clusters appear in this space too. 

\subsection {Ablation study: impact of CSVAE}

To investigate the role of CSVAE in the above architecture, we remove it and directly apply the mean and standard deviation of pretrained speaker embeddings to the extended reparametrization trick of Eq. \ref{extended_param_trick}. In this configuration, the latent variables are still shifted and the sampling is still done from the posterior given the condition. However the shift follows directions that have been pretrained to minimise loss of a different task, instead of directions that have been learned jointly with the rest of our system's parameters as when CSVAE is used. 

\begin{figure}[ht]
\centering
\vspace{-0.5cm}
\includegraphics[width=0.7\linewidth]{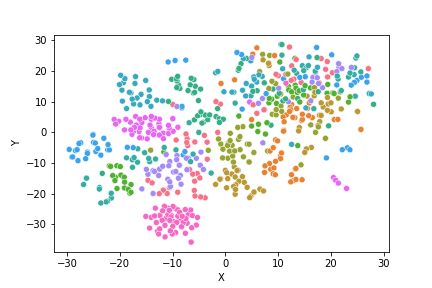}
\caption{t-SNE visualisation of the CPVAE latent space in the LCPVAE without CSVAE model. We directly apply mean and standard deviation of pretrained speaker embeddings as prior of CPVAE.}
\label{fig:cpvae-nocsvae-tsne}\end{figure}

In this case, some speaker clustering appears in the CPVAE space as can be seen in Figure \ref{fig:cpvae-nocsvae-tsne}. However, the clusters do not appear to be as well separated as in Figure \ref{fig:visualisations-lcpvae}. We also conducted a preference test in Amazon Mechanical Turk comparing this system with the baseline CVAE and following the same evaluation configuration as the one described in Section \ref{sec:eval}. The $p$-value of the binomial test comparing both models is greater than $p > \alpha=0.05$. Since the null hypothesis cannot be rejected, we conclude that there is no statistically significant difference between the LCPVAE without CSVAE model and the CVAE model. This result as well as the visualisations reinforce the advantage that the CSVAE provides, where speaker embeddings are trained specifically for the TTS task. The CSVAE could also be investigated as a speaker embeddings generator for external tasks, e.g. speaker verification, but this is left for future work. 

\vspace{-0.7cm}

\section{Conclusions}

In this paper, we presented a novel approach to sample from the VAE latent space. Using a hierarchical VAE structure, we learn a posterior conditioned on a signal, in our case speaker vector, and apply it as prior while sampling. This informative prior allows us to draw samples of varying prosody, while we gain controllability over the resulting disentangled latent space. In the multi-speaker scenario, speaker clusters appear in the latent space and samples are drawn for each speaker from its corresponding cluster. Moreover, there is no need for external speaker embeddings for the conditioning as they are generated during training. The proposed approach shows a significant improvement over TTS methods that use conditional VAEs.

In the herein study, the known condition is the speaker vector. We do, however, speculate that the proposed architecture can be applied to different conditions that influence speech, from language and speaking style, to more expressive ones such as emotion or even semantically informed conditions such as intent. Conditions more related to acoustics such as speaking rate could also be explored. Such an approach can find application to different branches of TTS, from voice conversion and data reduction to multi-lingual and multi-style modelling.  

\section{Acknowledgements}
The authors would like to thank Tom Merritt for his help with the data preparation, and Zack Hodari for insightful conversations on VAE informative priors. 

\bibliographystyle{IEEEtran}

\bibliography{mybib}

\end{document}